\documentclass[reprint,secnumarabic,amssymb, nobibnotes, aps, prd, onecolumn]{revtex4-2}
\usepackage{graphicx}
\usepackage{amsmath}
\usepackage{epstopdf}
\usepackage{setspace}

\begin{document}

\title{Transmission of waves and particles through the interface: reversibility and coherence}

\author{A.\,P. Meilakhs} 
\email[A.\,P. Meilakhs: ]{iestfi@gmail.com}
\affiliation{$^1$Ioffe Institute, 26 Politekhnicheskaya, St. Petersburg 194021, Russian Federation\\
 $^2$Centro Atomico Constituyentes, CNEA, Av. Gral. Paz 1499, San Martin, Buenos Aires, 1650, Argentina }
\date{\today}

\begin{abstract}
We examine the transmission of quantum particles (phonons, electrons, and photons) across interfaces, identifying universal patterns in diverse physical scenarios. Starting with classical wave equations, we quantize them and derive kinetic equations. Those are matching conditions for the distribution functions of particles at the interface. We note the time irreversibility of the derived kinetic equations — an essential feature for accurately describing irreversible processes like heat transport. We identify the juncture in our derivation where the time symmetry of wave equations is disrupted, it is the assumption of the non-coherence of incident waves. Consequently, we infer that non-coherent transmission through the interface exhibits time irreversibility. We propose an experiment to validate this hypothesis.
\end{abstract}

\maketitle

\section{Introduction}

Systematic investigations into transport phenomena associated with particle transmission through interfaces can be traced back to the seminal work by Landauer \cite{Landauer}, which describes one-dimensional electron transport in a disordered media.  Subsequently, the approach was expanded to encompass three dimensions in the manuscript \cite{Electrons1}. 

A pivotal phenomenon characterizing interfacial transport is. This phenomenon manifests as a temperature jump at the interface when heat  flows through the interface between media. A proportionality coefficient between the heat flux and the temperature jump is called Kapitza conductance \cite{Kap}. After the discovery of the phenomena, it was very soon realized that the temperature jump is due to the reflection of phonons at the interface \cite{Khal}. 

The science of Kapitza conductance is currently a prolific area of research. Some manuscripts delve into the dynamics of the crystal lattice at the interface through analytical approaches \cite{AnDin1,AnDin2,AnDin3} or computer simulations \cite{Din1,Din2,Din3,Din4,DinDMM,Din5}. Others study phonon kinetics at the interface with Boltzmann theory \cite{PhonKin1,PhonKin2,PhonKin3,PhonKin4} or the nonequilibrium Greens function method \cite{PhonGreen1,PhonGreen2,PhonGreen3}.  Additionally, ongoing experiments contribute to the landscape of research  \cite{Exp1,Exp2,Exp3,Exp4}. 

While the Kapitza jump phenomenon is predominantly linked with phonon transport through interfaces, the theory of electron transport has also undergone extensive development  \cite{Electrons7,Electrons8,Electrons2,Electrons3,Electrons4,Electrons5,Electrons6}. In the paper \cite{SemiKapitza} it was proposed to broaden the application of the Kapitza jump concept to encompass electron transport through interfaces.

The central framework in the theory of transport through interfaces is encapsulated by the Landauer-Buttiker formula \cite{Bruus, Buttiker}. However, this formula faces fundamental challenges. Initial applications, as demonstrated in Ref. \cite{Can}, reveal that a formal utilization of the Landauer-Buttiker formula for calculating the Kapitza resistance of a crystallographic plane in an ideal homogeneous crystal yields a non-zero result. This outcome contradicts expectations, as the correct calculation for such an imaginary boundary—where the reflection of phonons does not occur—should be precisely zero. The resolution to this paradox was first presented in Ref.,\cite{Maas}, where a one-dimensional model involving two bound harmonic strings was employed. In this case, the Landauer-Buttiker formula implies that the conductivity of such a system is proportional to the probability of particle transmission through the interface, denoted as $t$. However, considering the non-equilibrium state of incident phonons at the interface reveals that the conductivity is actually proportional to $t/(1-t)$. Notably, Landauer himself employed the expression $t/(1-t)$ in his original manuscript \cite{Landauer}.

Extending the arguments presented in the manuscript \cite{Maas} to more realistic three-dimensional models posed challenges. To elucidate and expand upon the concepts put forth in \cite{Maas}, the paper \cite{Me} introduced the notion of matching equations for the distribution function of phonons at the interface.  This approach permitted the description of non-equilibrium distribution functions of particles at the interface in a three-dimensional case.  The realization of this task, detailed in the paper \cite{KinDMM}, led to results that resolved the paradox highlighted in Ref. \cite{Can}:  it yielded a Kapitza resistance of zero for a crystallographic plane in a homogeneous crystal. Additionally, this method enabled calculations for both electron and phonon transport in a similar fashion, as demonstrated in \cite{SemiKapitza}, aligning with how Boltzmann equations uniformly apply to both electrons and phonons.

The first attempt to derive the matching equations for distribution functions was presented in Ref. \cite{Me}, concurrently with the introduction of the concept. In this paper, we aim to present a more rigorous derivation that places the resulting formula within the broader framework of quantum theory.  Section 2 serves as a preparatory groundwork, and Section 3 is dedicated to deriving the kinetic equation for phonon transport through the interface. In this context, the term "kinetic equation" refers to the equation governing one-particle distribution functions. Moving forward, Section 4 extends this derivation to electrons and explores its consequences.

It is crucial to emphasize that the application of the matching equations is limited to scenarios where particles approach the interface from infinity, or, at least, the distance between the interfaces in nanostructures is substantially larger than the typical particle wavelength and mean free path. In cases diverging from this condition, the interface alters the wave energy levels of particles \cite{Bastard1, Bastard2, Bastard3}, leading to distinct effects beyond the scope of our current discussion. These effects have been extensively explored, particularly for electrons \cite{SupLat3, SupLat4}, and phonons \cite{SupLat1,SupLat2}.

In our analysis, we exclusively focus on the interface's impact on kinetics. Section 5 of this study highlights the most fundamental property of the derived matching equations: their irreversibility. It is well-established that nonequilibrium processes, in general, exhibit irreversibility. Also, it is well known that the Boltzmann kinetic equation is irreversible. Consequently, it is natural that kinetic equations characterizing interfaces also possess irreversibility.  However, matching equations are much easier than Boltzmann equations: they are linear and they do not contain time derivatives. So it is much easier to find which assumption leads to the breaking of time symmetry.  In our investigation, we pinpoint this assumption as the incoherence of transmission through the interface. The meaning of incoherent transmission is carefully specified in Section 5.

 In Section 6 we come to the classic problem of light beams that go through the interface between media with different refraction indices. We propose an experiment that can demonstrate that coherent light transmits reversibly, while the transmission of noncoherent light is irreversible. We hope it can shed some light on the fundamental problem of an arrow of time, as it is stated in common statistical mechanics and thermodynamics textbooks \cite{LdStat, Prig}: how is it possible that macroscopic processes are irreversible, while microscopic processes are time-symmetric?

The concept of non-coherence is well-established in quantum theory \cite{DecohBook, Entanglement}, and the differentiation between coherent and noncoherent transmission has been a part of the discussion on transport properties related to interfaces \cite{Coherent1, Coherent2}. However, to the best of our knowledge, the presented manuscript suggests the explicit connection between coherence and reversibility for the first time.

\section{Classic atomic chain with interface}

We start our investigation with a one-dimensional classic atomic chain with the interface. Despite the extensive examination of this model in prior studies \cite{AnDin2, First}, we revisit this simple model to introduce a new notation and derive a formula that will prove instrumental in the subsequent paragraphs.

\begin{figure}
\includegraphics[width=0.48\textwidth]{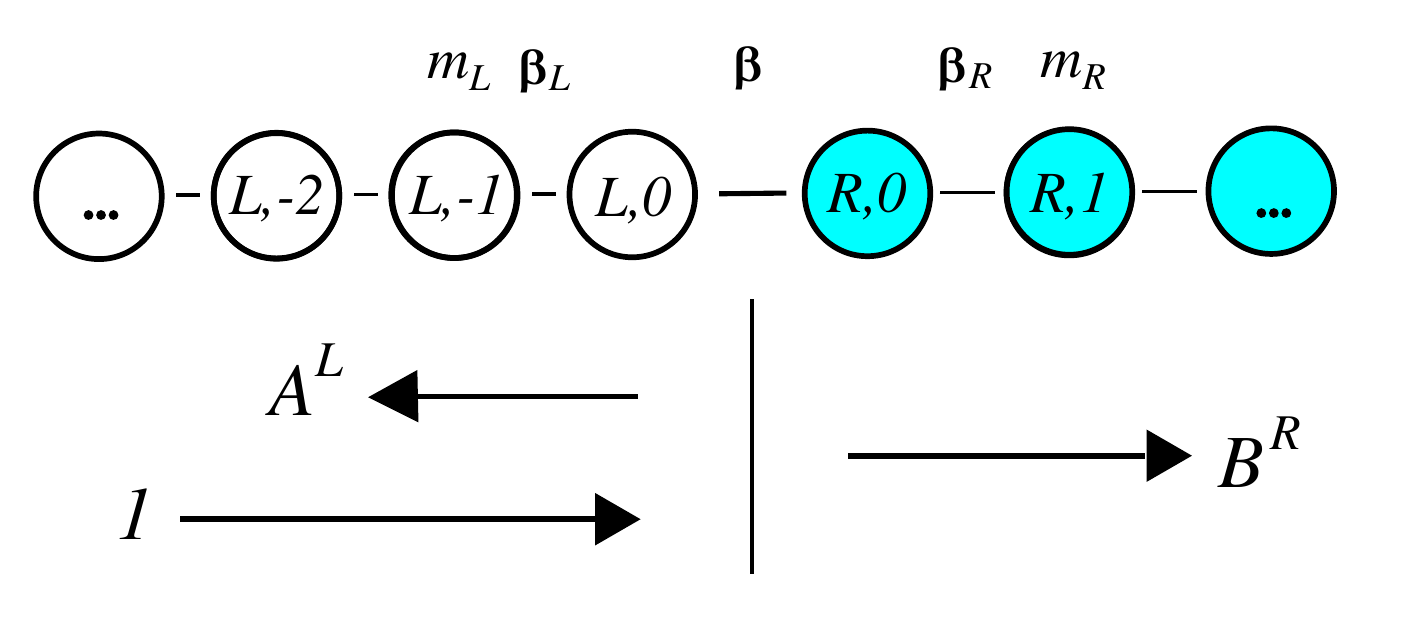}
\caption{Propagation of vibrations over two coupled semi-infinite one-dimensional chains. Indices $L$ and $R$ label the semi-infinite media on the left and right, respectively. The atoms in the left chain are numbered from $-\infty$ to $0$, and in the right chain from $0$ to $+\infty$. $\beta_{L,R}$ and $m_{L,R}$ denote the chain coupling parameters and atomic masses in the left and right media, respectively. The coupling parameter at the interface, specified by the vertical line, is denoted as $\beta$. The left medium encompasses incident and reflected waves and the right media encompasses the transmitted wave. The incident wave is assumed to have a unit amplitude, and $A_L$ and $B_R$ represent the amplitudes of the reflected and transmitted waves, respectively. }
\label{fig1}
\end{figure}

We consider a coupled one-dimensional chain characterized by a set of elastic constants $\beta^L, \beta^R, \beta$, which define interaction among atoms inside the left (L) and the right (R) media and at the interface. The masses of atoms in the media are $m^L$ and $m^R$, and interatomic distances are $a^L$ and $a^R$ (Fig. \ref{fig1}). The atoms in the left chain are numbered from $-\infty$ to $0$, and in the right chain from $0$ to $+\infty$. We denote the displacement of the atom on the left side as $u^L_n$ and on the right side as $u^R_n$. For all the atoms but those on the interface ($u^L_0$ and $u^R_0$) we write down the motion equation in the form
\begin{equation}
\ddot u_n = -\beta (u_n - u_{n-1}) -\beta (u_n - u_{n+1}),
\label{1first}
\end{equation}
where we omitted the side index $L$ or $R$, since the equation is the same for both indexes.

As it is well known \cite{Anselm}, the solution to such equation is (line denotes the complex conjugation)
\begin{equation}
u_n = A e^{i (\omega t \pm q a n)} +  \overline{A} e^{-i (\omega t \pm q a n)},
\label{1firstsol}
\end{equation}
while wave vector $q$ and frequency $\omega$ satisfies the dispersion relation. Starting from here we will only write down the first part and not the complex conjugate part. We substitute Eq. \eqref{1firstsol} into Eq. \eqref{1first}, and obtain
\begin{equation}
\omega^2 =  -\beta (1 - e^{i q^L a^L}) -\beta (1 - e^{-i q^L a^L}).
\label{1subs}
\end{equation} 
We simplify this expression to derive the dispersion relation
\begin{equation}
\omega = 2\sqrt{\beta/m} |\sin (qa/2)|.
\label{1dispersion}
\end{equation} 

We write down the motion equation for the atoms at the very interface
 \begin{align}
\ddot u^L_0 = -\beta^L (u^L_0 - u^L_{-1}) - \beta (u^L_0 - u^R_0) \nonumber \\
\ddot u^R_0 = -\beta (u^R_0 - u^L_0) - \beta^R (u^R_0 - u^R_1)
\label{1starteq}
\end{align}
 We want to seek the solution of this equation in the form of a superposition of plane waves (Eq. \ref{1firstsol}) with dispersion given by Eq. \ref{1dispersion}. The solution in such form will automatically satisfy motion equations for all atoms (Eq. \ref{1first}) but those interfacial ones (Eq. \ref{1dispersion}).

Let us assume the wave with a unit amplitude is incident on the interface from the left. Part of it is reflected and part of it is transmitted. We denote the amplitude of the reflected wave as $A^L$ and the amplitude of the transmitted wave as $B^R$. So the displacement of the $n$-th atom is
 \begin{align}
u^L_n &= e^{i (\omega t + q^L a^L n)} + A^L e^{i (\omega t - q^L a^L n)}
\nonumber \\
u^R_n &= B^R e^{i (\omega t + q^R a^R n)}.
\end{align}
 We substitute such superposition into equation \eqref{1starteq} and obtain 
 \begin{align}
\omega^2 ( 1 + A ) &= -\beta^L (1 + A^L - e^{i q^L a^L} - A^L e^{- i q^L a^L}) - \beta (1 + A^L - B^R)
 \nonumber \\
\omega^2 B &= -\beta (B^R - 1 - A^L) - \beta^R (B^R - B^R e^{i q a}).
\end{align}

To simplify this, we use Eq. \eqref{1subs} and obtain 
 \begin{align}
\beta^L (1 + A^L - e^{-i q^L a^L} - A^L e^{i q^L a^L}) &=  \beta (1 + A^L - B^R)
 \nonumber \\
\beta^R (B^R - B^R e^{i q^R a^R}) &= \beta (B^R - 1 - A^L) .
\end{align}
We transfer all unknown amplitudes to the left side of the equations
 \begin{align}
[\beta^L - \beta^L e^{ i q^L a^L}  - \beta] A^L + \beta B^R  &=
 -[\beta^L - \beta^L e^{-i q^L a^L} - \beta]
 \nonumber \\
\beta A^L + [\beta^R  - \beta^R e^{i q^R a^R} - \beta] B^R &= -\beta .
\label{1leftinc}
\end{align}

Now we introduce the following notation (based on Ref. \cite{AnDin3})
\begin{align}
M^L_\pm = \beta^L - \beta^L e^{\pm i q^L a^L}  - \beta
 \nonumber \\
M^R_\pm = \beta^R - \beta^R e^{\pm i q^R a^R}  - \beta
\label{1Ms}
\end{align}

With this notation, we put the system of equations (\ref{1leftinc}) into a matrix form
\begin{equation}
\begin{pmatrix} M^L_+ & \beta \\ \beta & M^R_+ \end{pmatrix}
\begin{pmatrix} A^L \\ B^R  \end{pmatrix} = 
- \begin{pmatrix} M^L_- \\ \beta \end{pmatrix}
\end{equation} 

We can obtain a similar set of equations in the case of the incidence of a wave with a unit amplitude on the interface from the right. We denote the amplitude of the reflected wave as $A^R$ and the amplitude of the transmitted wave as $B^L$. With all the same substitutions that we have done to obtain Eq. \eqref{1leftinc} we get
 \begin{align}
[\beta^L - \beta^L e^{ i q^L a^L}  - \beta] B^L + \beta A^R  &=  -\beta
 \nonumber \\
\beta B^L + [\beta^R  - \beta^R e^{i q^R a^R} - \beta] A^R &=  - [\beta^R  - \beta^R e^{-i q^R a^R} - \beta].
\label{1rightinc}
\end{align}

For our purposes, it is convenient  to combine both systems of equations (\ref{1leftinc}, \ref{1rightinc}) into one matrix equation
\begin{equation}
\begin{pmatrix} M^L_+ & \beta \\ \beta & M^R_+ \end{pmatrix}
\begin{pmatrix} A^L & B^L \\ B^R & A^R \end{pmatrix} = 
- \begin{pmatrix} M^L_- & \beta \\ \beta & M^R_- \end{pmatrix}
\begin{pmatrix} 1 & 0 \\ 0 & 1 \end{pmatrix}
\label{1maineq}
\end{equation} 

We would like to introduce a compact notation for the two-by-two matrices used:
\begin{equation}
\mathcal M_\pm = \begin{pmatrix} M^L_\pm & \beta \\ \beta & M^R_\pm \end{pmatrix}, \,
\mathcal A = \begin{pmatrix} A^L & B^L \\ B^R & A^R \end{pmatrix} 
\label{1mathcal}
\end{equation} 

Now we can write the solution of the equation \eqref{1maineq} compactly
\begin{align}
\mathcal A = -\mathcal M_+^{-1}\, \mathcal M_-.
\label{1solution}
\end{align} 

For explicit computations, we will need the inverse of $\mathcal M_+$. We use the formula:
\begin{equation}
 \mathcal M_+^{-1} = \frac{\mathrm{adj}\, \mathcal M_+}{\mathrm{det}\, \mathcal M_+} =
\frac{1}{M^L_+ M^R_+ -\beta^2} \begin{pmatrix} M^L_+ & -\beta \\ -\beta & M^R_+ \end{pmatrix},
\label{1inverse}
\end{equation} 
here $\mathrm{adj}\, \mathcal M_+$ is the adjoint matrix.

We want to investigate the properties of the matrix of amplitudes $\mathcal A$. For the solution to hold physical significance, it must adhere to the conservation of energy flow condition. Considering that the energy flow associated with a plane wave is proportional to the group velocity of the wave, the density of the media, and the square of amplitudes, the following relations arise:
\begin{align}
 \rho^L v^L |A^{L}|^2 + \rho^R v^R |B^{R}|^2 = \rho^L v^L
 \nonumber \\
 \rho^R v^R |A^{R}|^2 + \rho^L v^L |B^{L}|^2 = \rho^R v^R.
\label{1flowcons}
\end{align}
Energy flow is also proportional to frequency. However, given that we are dealing with a harmonic chain, the frequencies of all waves in the same equations are equal to each other and, therefore, cancel out.

We want to accompany these equalities with equality:
\begin{align}
 \rho^L v^L \overline{A^L} B^L +\rho^R v^R \overline{B^R} A^R =  0.
\label{1flowort}
\end{align}
Equations \eqref{1flowcons} and \eqref{1flowort} together constitute a condition resembling orthonormality of flows. The physical significance of the condition \eqref{1flowort} will be elucidated later.

It is convinient to incorporate both conditions (\ref{1flowcons}, \ref{1flowort}) into one matrix equation. We introduce the flow matrix
\begin{equation}
\mathcal F =  \begin{pmatrix} \rho^L v^L & 0 \\ 0 & \rho^R v^R \end{pmatrix},
\label{1flowmat}
\end{equation} 
which are coefficients that relate squares of amplitudes to flows.

In this notation we have (dagger denotes the Hermitian conjugation)
\begin{equation}
 \mathcal A^\dag \mathcal F  \mathcal A = \mathcal F.
\label{1mainres}
\end{equation} 
This matrix equation contains four ordinary equations, but one of them is a complex conjugate of equation \eqref{1flowort}.

Let's verify identities (\ref{1flowcons}, \ref{1flowort}) or, which is the same, the identity \eqref{1mainres}. We will start with the first equation of\eqref{1flowcons}. We would substitute the amplitudes $A^L, B^R$ in the form given by equations (\ref{1mathcal}, \ref{1solution}, \ref{1inverse}) and use the properties of introduced coefficients \eqref{1Ms} $\overline{M^{L,R}_\pm} = M^{L,R}_\mp$:
\begin{align}
\rho^L v^L (M^L_+ M^R_- -  \beta^2) (M^L_- M^R_+ -  \beta^2) &+
 \rho^R v^R \beta^2 (M^L_- - M^L_+)  (M^L_+ - M^L_-) =
\rho^L v^L (M^L_- M^R_- -  \beta^2) (M^L_+ M^R_+ -  \beta^2).
\label{1calc}
\end{align}
Here we have multiplied both sides on determinants of the matrices that arise from Eq. \eqref{1inverse}.

We expand the brackets in \eqref{1calc} and after canceling out the similar terms on both sides we get
\begin{equation}
 \rho^L v^L (M^R_+ - M^R_-) + \rho^R v^R (M^L_- - M^L_+) = 0  .
\label{1Fin}
\end{equation} 
We use definition of M-s \eqref{1Ms} to get
\begin{equation}
 \rho^L v^L \beta^R \sin q^R a^R = \rho^R v^R \beta^L \sin q^L a^L .
\end{equation} 
But it is true, because for a one-dimensional harmonic chain $\rho v = \beta \sin qa$ \cite{Kotkin}. 

For the second equation in \eqref{1flowcons} the calculations are completely analogous. For the equation \eqref{1flowort} after substitutions of the amplitudes (\ref{1mathcal}, \ref{1solution}, \ref{1inverse}) we have
\begin{align}
\rho^L v^L (M^L_+ M^R_- -  \beta^2) \beta  (M^R_+ - M^R_-) &+
\rho^R v^R  \beta (M^L_- - M^L_+) (M^L_+ M^R_- -  \beta^2) = 0.
\end{align} 
Here we have already canceled the similar denominator for both summands. After dividing by similar terms, we obtain exactly the Eq. \eqref{1Fin} which we have already verified.

Thus we have proven the identity \eqref{1mainres}, which is the main result of this section. We will heavily rely on it in the next sections.

we aim to explore another aspect. In a prior study \cite{AnDin2}, it was demonstrated that in the low-frequency limit, the amplitudes of transmission and reflection become independent of the value of $\beta$, the constant of interaction between the chains. The amplitudes equal to those found for two connected elastic strings:
\begin{align}
\mathcal A =
\setstretch{1.5}
 \frac{1}{\rho^L v^L + \rho^R v^R}
 \begin{pmatrix} \rho^L v^L - \rho^R v^R & 2 \rho^R v^R \\
2 \rho^L v^L & \rho^R v^R - \rho^L v^L \end{pmatrix}.
\label{1elastic}
\end{align} 
We see that it is also independent of the frequency $\omega$. With imminent use in mind, we want to see, which properties of coefficients in Eq. \eqref{1maineq} make it this way. 

 In the limit $q \rightarrow 0$, formulae \eqref{1Ms} can be rewritten as
\begin{align}
M^L_\pm = \mp i \omega \beta^L a^L/v^L - \beta
 \nonumber \\
M^R_\pm = \mp i \omega \beta^R a^R/v^R - \beta.
\label{1longwave}
\end{align}
 We substitute these expressions into definition \eqref{1mathcal} and we see that the matrices $\mathcal M$ can be presented as a sum of two components:
\begin{align}
\mathcal M_\pm = \beta \begin{pmatrix} - 1 & 1 \\ 1 & - 1 \end{pmatrix} \mp
\omega \begin{pmatrix}  i \beta^L a^L/v^L  & 0 \\0 &  i \beta^R a^R/v^R  \end{pmatrix}.
\label{1components}
\end{align} 
The first summand is singular and is the same for both matrices $\mathcal M_+$ and $\mathcal M_-$. The second is proportional to the small parameter $\omega$.

We recall the formula for amplitudes \eqref{1solution} and the formula for inverse matrix \eqref{1inverse}. For a singular matrix $\mathcal S$ the product $(\mathrm{adj}\, \mathcal S)\, \mathcal S = 0$. In our case that means
\begin{align}
\begin{pmatrix}  1 & 1 \\ 1 & 1 \end{pmatrix} \begin{pmatrix}  1 & -1 \\ -1 & 1 \end{pmatrix} = 0,
\end{align} 
which is easy to check. We also discard terms that are proportional to $\omega^2$. So we have
\begin{align}
(\mathrm{adj}\,\mathcal M_+) \mathcal M_- = i  \omega \beta \left(\begin{pmatrix} \beta^R a^R/v^R  & 0 \\0 &  \beta^L a^L/v^L  \end{pmatrix} \begin{pmatrix}  1 & -1 \\ -1 & 1 \end{pmatrix} -
\begin{pmatrix}  1 & 1 \\ 1 & 1 \end{pmatrix}
\begin{pmatrix} \beta^L a^L/v^L  & 0 \\0 & \beta^R a^R/v^R \end{pmatrix} \right).
\label{1partial}
\end{align} 
We see that it is proportional to $\omega \beta$. 

Concerning the determinant of $\mathcal M_+$, from formula \eqref{1inverse}, the determinant of a singular matrix is zero, and we also want to discard the second order in $\omega$ terms, so the determinant is also proportional  $\omega \beta$. Expression for the amplitude matrix is \eqref{1partial} divided by determinant of  $\mathcal M_+$. So $\omega \beta$ cancels out.

To verify our reasoning, we perform the direct computation, descard terms proportional to $\omega^2$ and obtain
\begin{align}
\mathcal A =
\setstretch{1.5}
 \frac{1}{\beta^L a^L/v^L +  \beta^R a^R/v^R}
\begin{pmatrix} \beta^L a^L/v^L -  \beta^R a^R/v^R & 2 \beta^R a^R/v^R \\
2 \beta^L a^L/v^L &  \beta^R a^R/v^R -\beta^L a^L/v^L \end{pmatrix}.
\end{align} 
We use expressions $v=a\sqrt{\beta/m}$ and $\rho = m/a$ and with simple algebra we arrive to identity \eqref{1elastic}. 

In summary, the amplitudes at zero frequency are independent of both $\beta$ and $\omega$ because the $\mathcal M$ matrices are singular at zero frequency. The significance of the qualitative arguments presented above lies in their applicability to more complex systems beyond the one currently under discussion. For instance, in a study \cite{AnDin3}, the equations governing the transmission of lattice oscillations through the interface between three-dimensional crystals are explored. We observe that in that case as well, the $\mathcal M$ matrices are singular in the low-frequency limit. Consequently, following the presented arguments, the amplitudes of transmission and reflection become frequency-independent and do not vary with the value of $\beta$, the constant of inter-lattice interaction.

 The expression \eqref{1components} shows what exactly does it mean that $\omega$ is "small". To treat $\omega$ as a small parameter both conditions should be satisfied: $\omega a^L/v^L \ll \beta/\beta^L$ and $\omega a^R/v^R \ll \beta/\beta^R$. This conditions have a clear meaning when we pass from the frequency to the wavelengthes $\lambda^{L,R}$ in both media:
\begin{align}
2\pi \frac{a^L}{\lambda^L} \ll \frac{\beta}{\beta^L},
2\pi \frac{a^R}{\lambda^R} \ll \frac{\beta}{\beta^R}.
\end{align} 
As we can see, when the interfacial coupling parameter $\beta$ is of the same order of magnitude as the coupling parameters within the media, elastic continuum models work well for wavelengths up to the lattice parameters. This implies that continuum theory is a valid approximation to describe the transmission of waves with frequencies up to the Debye frequency of the media. This observation provides an explanation for the surprisingly accurate approximation of the temperature dependence of Kapitza resistance by the Debye function \cite{KinDMM}.

%%%%%%%%%%%%%%%%%%%%%%%%%%%%%%%%%%%%%%

\section{Quantization of atomic chain: transmission of phonons through the interface}

We commence with a quantization of a simple atomic chain without the interface. In most books, the suggested approach is one of the following. Either we solve the classical problem and then interchange amplitudes of the normal modes of the chain to the annihilation and creation operators \cite{Davydov} or we write down the Hamiltonian and then diagonalize it with the Bogoliubov transform \cite{LevitovShitov}. However, we opt for a different method based on the Heisenberg representation, as it aligns well with our subsequent exploration of the coupled chain problem.

The Hamiltonian for a simple atomic chain can be written as \cite{LevitovShitov}
\begin{equation}
\hat H = \sum_i \left(\frac{\hat p_i^2}{2m} + \frac{\beta (\hat u_i - \hat u_{i-1})^2}{2} \right),
\label{2Hamilton}
\end{equation} 
where $\hat u_i$ and $\hat p_i$ are operators of displacement and momentum of $i$-th atom. Those operators obey the standard coordinate-momentum commutation relations:
\begin{equation}
[\hat u_j, \hat p_i] = i \hbar \, \delta_{ij}.
\label{2Heisenberg}
\end{equation} 

With this commutation relations, we can deduce
\begin{align}
[\hat u_i^2, \hat p_i] = u_i^2 p_i - p_i u_i^2 = u_i  u_i p_i - u_i  p_i u_i -  i \hbar  u_i = u_i [\hat u_j, \hat p_i]  -  i \hbar  u_i =  - 2  i \hbar \,  u_i.
\label{2H1}
\end{align} 
And similarly 
\begin{align}
[\hat p_i^2, \hat u_i] =  - 2  i \hbar \, p_i.
\label{2H2}
\end{align} 

We write down Heizenberg relations for displacements and momenta operators. With the hamiltonian \eqref{2Hamilton} and the help of Eqs. (\ref{2H1}, \ref{2H2}) we obtain
\begin{align}
\dot {\hat p}_i = \frac{i}{\hbar} [\hat H, \hat p_i] &= - \beta (\hat u_i - \hat u_{i-1})
\nonumber \\
\dot {\hat u}_i = \frac{i}{\hbar} [\hat H, \hat u_i] &=  \hat p_i/m.
\end{align} 

These equations are precisely equivalent to Hamilton's equations for classic chain and therefore have the same solution. We differentiate with respect to time the second one of them and substitute the first one into the obtained equation. The result is
\begin{equation}
\ddot{ \hat u}_n = -\beta (\hat u_n - \hat u_{n-1}) -\beta (\hat u_n - \hat u_{n+1}),
\label{2Eq}
\end{equation}
and if we substitute
\begin{equation}
\hat u_n = \sum_\omega \left( \hat A e^{i (\omega t \pm q a n)} +
 \hat A^\dag e^{- i (\omega t \pm q a n)} \right).
\label{2Sol}
\end{equation}
we obtain the exact identity, as long as dispersion relation \eqref{1dispersion} holds.

For the chain with the interface, Hamiltonian is
\begin{align}
\hat H = \sum_{i = -\infty}^0 \left(\frac{\hat p_i^{L^2}}{2m} + \frac{\beta (\hat u_i^L - 
\hat u_{i-1}^L)^2}{2} \right) +
 \frac{\beta (\hat u_0^L - \hat u_0^R)^2}{2} +  \sum_{i = 1}^{\infty} \left(\frac{\hat p_i^{R^2}}{2m} + \frac{\beta (\hat u_i^R - \hat u_{i-1}^R)^2}{2} \right),
\label{2Hamilton2}
\end{align} 

We write down Heisenberg equations for interfacial atoms on both sides of the interface and obtain
 \begin{align}
\ddot {\hat u}^L_0 = -\beta^L (\hat u^L_0 - \hat u^L_{-1}) - \beta (\hat u^L_0 - \hat u^R_0) \nonumber \\
\ddot {\hat u}^R_0 = -\beta (\hat u^R_0 - \hat u^L_0) - \beta^R (\hat u^R_0 - \hat u^R_1).
\label{2System}
\end{align}
Up to this point, our calculations closely mirror those performed for a classical chain. However, a pivotal distinction emerges between quantum and classical cases at this juncture: the amplitudes are operators. They act on the entire state of the system, and we can no longer assume unit amplitude for one of them and zero amplitude for most of the rest. Instead, we must consider and retain all amplitudes in our calculations.

Phonons that are incident on the interface and those departing from the interface should be treated separately. To distinguish their directions of propagation, we introduce arrow indices $\leftarrow$ and $\rightarrow$ (indicating the wave travels from the right to the left and vice versa, respectively). Employing this notation, we express the displacement operators for the atoms on the left and right sides of the interface:
\begin{align}
\hat u_n^L = \sum_\omega \left( \hat A^L_{\leftarrow} e^{i (\omega t - q a n)}
 + \hat A^L_{\rightarrow} e^{i (\omega t + q a n)} \right) + H.C.  \nonumber \\
\hat u_n^R = \sum_\omega \left( \hat A^R_{\leftarrow} e^{i (\omega t - q a n)}
 + \hat A^R_{\rightarrow} e^{i (\omega t + q a n)} \right)  + H.C.
\label{2fullsum}
\end{align}
H.C. stands for the Hermitian conjugate. Modes that are incident on the interface now have indices ($L, \rightarrow$) and ($R, \leftarrow$) and the modes that depart from the interface have indices ($L, \leftarrow$) and ($R, \rightarrow$).

We formally introduce the displacement operators of "virtual atoms", atom number one in the left crystal and atom number minus one in the right crystal
\begin{align}
\hat u_1^L = \sum_\omega \left( \hat A^L_{\leftarrow} e^{i (\omega t - q a)}
 + \hat A^L_{\rightarrow} e^{i (\omega t + q a )} \right) + H.C.  \nonumber \\
\hat u_{-1}^R = \sum_\omega \left( \hat A^R_{\leftarrow} e^{i (\omega t + q a)}
 + \hat A^R_{\rightarrow} e^{i (\omega t - q a)} \right)  + H.C.
\end{align}
We substitute $\hat u_1^L$ and $\hat u_{-1}^R$ into Eq. \eqref{2Eq} for the zeroth atoms on both sides, then substitute the result into \eqref{2System}:
 \begin{align}
\beta^L (\hat u^L_0 - \hat u^L_1) = \beta (\hat u^L_0 - \hat u^R_0) \nonumber \\
\beta^R (\hat u^R_0 - \hat u^R_{-1}) = \beta (\hat u^R_0 - \hat u^L_0) .
\label{2System2}
\end{align}
This way we excluded time derivatives from the equations. We could have applied the exact same trick for the classical chain to simplify Eq. \eqref{1starteq}, however here it is really useful because it saves a lot of effort.

We substitute definition \eqref{2fullsum} to the system of equations \eqref{2System2}. We multiply equations by $e^{i \omega t }$ and integrate over time from minus infinity to infinity, so only one frequency is left in the system of equations.  This way we obtain 
\begin{align}
\beta^L (1- e^{- i q^L a^L}) \hat A^L_{\leftarrow} +
 \beta^L (1- e^{ i q^L a^L}) \hat A^L_{\rightarrow} =
\beta (\hat A^L_{\leftarrow} + \hat A^L_{\rightarrow} -  \hat A^R_{\leftarrow} - \hat A^R_{\rightarrow})
\nonumber \\
\beta^R (1- e^{ i q^R a^R}) \hat A^R_{\leftarrow} +
 \beta^R (1- e^{- i q^R a^R}) \hat A^R_{\rightarrow} =
\beta (\hat A^R_{\leftarrow} + \hat A^R_{\rightarrow} - \hat A^L_{\leftarrow} - \hat A^L_{\rightarrow}).
\end{align} 
After multiplying equations by $e^{i \omega t }$ and integrating over time we also got rid of Hermitian conjugate parts of $\hat u$ operators. To obtain the equation for Hermitian conjugated amplitude operators we need to multiply by $e^{-i \omega t }$ before integrating. We will proceed to write equations for amplitudes keeping in mind that there is a symmetric equation for the Hermitian conjugated amplitudes.

We group the amplitudes of departing waves on one side and the amplitudes of the incident waves on the other side. We use the notation \eqref{1Ms} and thus we arrive to
\begin{align}
M^L_+ \hat A^L_{\leftarrow} + \beta \hat A^R_{\rightarrow} = 
- M^L_-  \hat A^L_{\rightarrow} - \beta \hat A^R_{\leftarrow}
\nonumber \\
\beta \hat A^L_{\leftarrow} + M^R_+ \hat A^R_{\rightarrow} =
-\beta \hat A^L_{\rightarrow} - M^R_- \hat A^R_{\leftarrow}.
\label{2maineq}
\end{align} 

We rewrite this equation in the matrix form
\begin{align}
\mathcal M_+\begin{pmatrix} \hat A^L_{\leftarrow} \\ \hat A^R_{\rightarrow} \end{pmatrix} = 
-\mathcal M_-\begin{pmatrix} \hat A^L_{\rightarrow } \\ \hat A^R_{\leftarrow} \end{pmatrix}.
\label{2amps}
\end{align} 
By formula \eqref{1solution} it follows that
\begin{align}
\begin{pmatrix} \hat A^L_{\leftarrow} \\ \hat A^R_{\rightarrow} \end{pmatrix} = 
\mathcal A\begin{pmatrix} \hat A^L_{\rightarrow } \\ \hat A^R_{\leftarrow} \end{pmatrix}.
\label{2amps}
\end{align} 

Our objective is to transition from describing the system in terms of amplitudes to describing it in terms of the occupation numbers of particles. We frame the quantum problem of phonon transmission through the interface as determining the probabilities of transition from the initial state, specified by the occupation numbers of phonons incident on the interface, to the final state, specified by the occupation numbers of phonons departing from the interface.

To achieve this, we aim to convert the relation between amplitude operators to the relation between phonon annihilation operators (and, consequently, hermitian conjugate amplitude operators to phonon creation operators). We leverage the fact that the amplitude operators are proportional to annihilation operators \cite{Anselm}: $\hat A^L = \nu^L \hat a^L, \hat A^R = \nu^R \hat a^R$, where $\nu^L$ and $\nu^R$ are normalization constants. We omit the arrow indices, as both directions in each crystal are symmetric, and the normalization constants are the same. Our goal is to determine the values of $\nu^L$ and $\nu^R$.

 We can introduce a normalization matrix
\begin{align}
\mathcal N = \begin{pmatrix} \nu^L & 0 \\ 0 & \nu^R \end{pmatrix},
\label{2norms}
\end{align} 
 and write down 
\begin{align}
\mathcal N \begin{pmatrix} \hat a^L_{\leftarrow} \\ \hat a^R_{\rightarrow} \end{pmatrix} = 
\mathcal A \mathcal N \begin{pmatrix} \hat a^L_{\rightarrow } \\ \hat a^R_{\leftarrow} \end{pmatrix}.
\end{align} 
whence immediately follow
\begin{align}
\begin{pmatrix} \hat a^L_{\leftarrow} \\ \hat a^R_{\rightarrow} \end{pmatrix} = 
\mathcal N^{-1} \mathcal A \mathcal N \begin{pmatrix} \hat a^L_{\rightarrow } \\ \hat a^R_{\leftarrow} \end{pmatrix} = \mathcal U \begin{pmatrix} \hat a^L_{\rightarrow } \\ \hat a^R_{\leftarrow} \end{pmatrix}.
\label{2connect}
\end{align} 
Here the last equality defines the new matrix $\mathcal U$. We denote it this way because this matrix will turn out to be unitary. But at this point, we do not know it yet. We proceed to prove the unitarity of
$\mathcal U$.

We will rewrite the last equation explicitely, through elements of matrix  $\mathcal U$
\begin{align}
 \hat a^L_{\leftarrow} = u_{11} \hat a^L_{\rightarrow} + u_{12} a^R_{\leftarrow}
\nonumber \\
 \hat a^R_{\rightarrow} = u_{21} \hat a^L_{\rightarrow} + u_{22} a^R_{\leftarrow}.
\label{2hats}
\end{align} 

Creation and annihilation operators taken at the same time follow well-known \cite{LevitovShitov} commutation relation. For phonons incident on the interface we have
\begin{align}
[\hat a^{L^\dag}_{\rightarrow}, a^L_{\rightarrow}] =
 [\hat a^{R^\dag}_{\leftarrow}, a^R_{\leftarrow}] = 1
 \nonumber \\
[\hat a^{L^\dag}_{\rightarrow}, a^R_{\leftarrow}] =  [\hat a^{R^\dag}_{\leftarrow}, a^L_{\rightarrow}] = 0.
\label{2comms}
\end{align} 
Analogous relations hold for creation and annihilation operators corresponding to phonons departing from the interface. We frame the state of departing phonons as a subsequent state of incident phonons. In general, the commutation relations for operators at different moments are unknown. However, in the case under investigation, the connection is specified by Eqs. (\ref{2connect}, \ref{2hats}). By commuting both sides of the system of equations \eqref{2hats} first with $\hat a^{L^\dag}{\rightarrow}$, then with $\hat a^{R^\dag}{\leftarrow}$, and using relations \eqref{2comms}, we can obtain the commutators:
\begin{align}
[\hat a^{L^\dag}_{\rightarrow}, \hat a^L_{\leftarrow}] &= u_{11} 
\nonumber \\
[\hat a^{R^\dag}_{\leftarrow}, \hat a^L_{\leftarrow}] &=  u_{12} 
\nonumber \\
[\hat a^{L^\dag}_{\rightarrow}, \hat a^R_{\rightarrow}] &= u_{21} 
\nonumber \\
[\hat a^{R^\dag}_{\leftarrow}, \hat a^R_{\rightarrow}] &= u_{22}.
\label{2inprod}
\end{align} 

To investigate the properties of the matrix $\mathcal U$ we coomute both sides of Eq. \eqref{2hats} and find
\begin{align}
[\hat a^{L^\dag}_{\leftarrow}, \hat a^L_{\leftarrow}] = [u_{11}^* \hat a^{L^\dag}_{\rightarrow} + u_{12}^* a^{R^\dag}_{\leftarrow}, u_{11} \hat a^L_{\rightarrow} + u_{12} a^R_{\leftarrow}] =
\nonumber \\
= |u_{11}|^2 [\hat a^{L^\dag}_{\rightarrow},  \hat a^L_{\rightarrow}] + u_{11}^* u_{12} [\hat a^{L^\dag}_{\rightarrow}, \hat a^R_{\leftarrow}] +  u_{12}^* u_{11} [\hat a^{R^\dag}_{\leftarrow}, \hat a^L_{\rightarrow}] + |u_{12}|^2
 [\hat a^{R^\dag}_{\leftarrow}, \hat a^R_{\leftarrow}] .
\end{align} 
With the help of Eq. \eqref{2comms} we find
\begin{align}
 |u_{11}|^2 + |u_{12}|^2 = 1
\end{align} 
Similarly
\begin{align}
[\hat a^{L^\dag}_{\leftarrow}, \hat a^R_{\rightarrow}] = u_{11}^* u_{21} + u_{12}^*u_{22} = 0 
\nonumber \\
[\hat a^{R^\dag}_{\leftarrow}, \hat a^R_{\leftarrow}] = |u_{21}|^2 + |u_{22}|^2 = 1.
\end{align} 
But these conditions mean the raws of matrix $\mathcal U$ form an orthonormal basis which implies $\mathcal U$ is unitary.

Throughout this calculation we could have seen that comutators behaves very similar to inner product in a complex vector space. The equations \eqref{2inprod} would be exactly the formula for the matrix of the change of a basis if brackets denoted the inner product. In this interpretation relations \eqref{2comms} are orthonormality conditions for bases.

So we know the relation \eqref{1mainres} for amplitude matrix and connection between amplitude matrix and some yet unknown unitary matrix \eqref{2connect}. We need to find such $\mathcal N$ that its substitution into \eqref{1mainres} in the form
\begin{align}
\mathcal A = \mathcal N \mathcal U \mathcal N^{-1}
\label{2transform}
\end{align}
will transform this relation into the relation that defines a unitary matrix. It is easy to see that the required matrix is $N = F^{-1/2}$.

Let us check it:
\begin{align}
 \mathcal F^{1/2} \mathcal U^\dag \mathcal F^{-1/2} \mathcal F  \mathcal F^{-1/2} \mathcal U \mathcal F^{1/2} = \mathcal F <=> \nonumber \\
<=> \mathcal F^{1/2} \mathcal U^\dag  \mathcal U \mathcal F^{1/2} = \mathcal F <=> \nonumber 
\\
\mathcal U^\dag  \mathcal U = \mathcal E, 
\label{2unitary}
\end{align} 
where $\mathcal E$ is identity matrix. But this is exactly the definition of matrix $\mathcal U$ being unitary. So the normalization constants are $(\rho^L v^L)^{-1/2}$ and $(\rho^R v^R)^{-1/2}$. Now we see why relation \eqref{1mainres} looks so similar to orthonormality condition. $\mathcal A$ turns out to be matrix of unitary transformation but in non-normalized basis.

 We want to note here briefly, that in an arbitrary basis, a matrix of unitary transformation, $\mathcal U$ obeys the relation $\mathcal U^\dag \mathcal H \mathcal U = \mathcal H$ where $\mathcal H$ is some hermitian matrix. $\mathcal H$ is a diagonal matrix only on an orthogonal basis and is an identity matrix on an orthonormal basis. Choosing the normalization constants corresponds to multiplying on diagonal matrix \eqref{2norms}. This means the transition from the amplitudes matrix $\mathcal A$ to the unitary transformation matrix is only possible if amplitudes satisfy the condition \eqref{1mainres}, and it is not possible for an arbitrary matrix of amplitudes. We have never postulated the formula \eqref{1mainres}, we posed the ordinary problem of classic mechanics and derived it from there. I think that gives us some confidence that we indeed have found some right point of view on this problem. Also, in quantum mechanics, the transformation from one moment of time to some moment in the future should be given by unitary transformation, because this type of transformation preserves the dot product and so it guarantees that the total probability of all the possible states of the system remains equal to one, because the probabilities are defined with the dot product. Now, in our treatment, we think about the incident phonons as those preceding the departing phonons. The fact the transformation between their amplitudes turns out to be the unitary transformation confirms the correctness of such a point of view.

Now we want to derive the transformation of distribution functions of phonons that are incident on the interface to those departing from the interface. We write down Eq. \eqref{2connect} explicitly, with $\mathcal U$ being defined by the inverse transform of \eqref{2transform}
\begin{align}
 \hat a^L_{\leftarrow} = A^L \hat a^L_{\rightarrow } + \sqrt{\frac{\rho^L v^L}{\rho^R v^R}} B^L \hat a^R_{\leftarrow}
\nonumber \\
 \hat a^R_{\rightarrow}  = \sqrt{\frac{\rho^R v^R}{\rho^L v^L}} B^R \hat a^L_{\rightarrow } + 
A^R \hat a^R_{\leftarrow}.
\label{2AmpOps}
\end{align} 
We recall that for all the equations for $\hat a$ there are Hermitian conjugated counterparts for $\hat a^\dag$. Using this we calculate the number of particles operators $ \hat n = \hat a^\dag \hat a $:
\begin{align}
 \hat n^L_{\leftarrow} = |A^L|^2 \hat n^L_{\rightarrow } + 
 \frac{\rho^L v^L}{\rho^R v^R} |B^L|^2 \hat n^R_{\leftarrow} + 
\sqrt{\frac{\rho^L v^L}{\rho^R v^R}}
\left( \overline A^L B^L \hat a^{L^\dag}_{\rightarrow } \hat a^R_{\leftarrow} +
 A^L \overline B^L \hat a^L_{\rightarrow } \hat a^{R^\dag}_{\leftarrow} \right)
\nonumber \\
\hat n^R_{\rightarrow} = \frac{\rho^R v^R}{\rho^L v^L} |B^R|^2 \hat n^L_{\rightarrow } + 
|A^R|^2 \hat n^R_{\leftarrow} +
  \sqrt{\frac{\rho^R v^R}{\rho^L v^L}} \left( 
\overline B^R A^R \hat a^{L^\dag}_{\rightarrow } \hat a^R_{\leftarrow} +
B^R \overline A^R \hat a^L_{\rightarrow } \hat a^{R^\dag}_{\leftarrow}
 \right).
\label{2NumbOps}
\end{align} 
We denote the state of the departing particles $| D \rangle$. If the state that creation and annihilation operators act on has a defenite number of particles, then
\begin{align}
\langle D| \hat a_i^\dag \hat a_j | D \rangle = n_i \delta_{ij}.
\label{2Kronecker}
\end{align} 
So if the state of the departing particles have a defenite number of particles (in Section 5 we will clarify this point further), the matrix elements of operators in the brackets of equation \eqref{2NumbOps} are zero and for ocupation numbers of particles we have
\begin{align}
 n^L_{\leftarrow} = |A^L|^2 n^L_{\rightarrow} + \frac{\rho^L v^L}{\rho^R v^R} |B^L|^2 n^R_{\leftarrow}
\nonumber \\
n^R_{\rightarrow} = \frac{\rho^R v^R}{\rho^L v^L} |B^R|^2 n^L_{\rightarrow } + 
|A^R|^2 n^R_{\leftarrow}.
\label{2PhononKinetic}
\end{align} 
It is exactly the equations that have been derived in Ref. \cite{Me}. But the method that we have introduced here brings much more insight since it reveals the underlying unitary transformation between annihilation operators  We will not investigate the derived equations here since it is already done in \cite{Me}, but we will do the investigation with analogous equations for electrons that we will derive in the next section.

%%%%%%%%%%%%%%%%%%%%%%%%%%

\section{Electrons transmission through the interface}
 There are many different Hamiltonians that describe electron states in a crystal. Mathematically the  one that produces the equations closest to the dynamics of crystal lattice equations is the tight binding Hamiltonian. The annihilation operators of electrons at one given atom correspond to the displacement of a given atom in the equations of lattice dynamics \eqref{2Eq}. Transferring the ideas of the previous paragraph into a tight-binding method framework would be very straightforward. Unfortunately, it seems impossible to write down the conservation of flow at the interface equation \eqref{1mainres} in the framework of the tight-binding model. 

\begin{figure}
\includegraphics[width=0.48\textwidth]{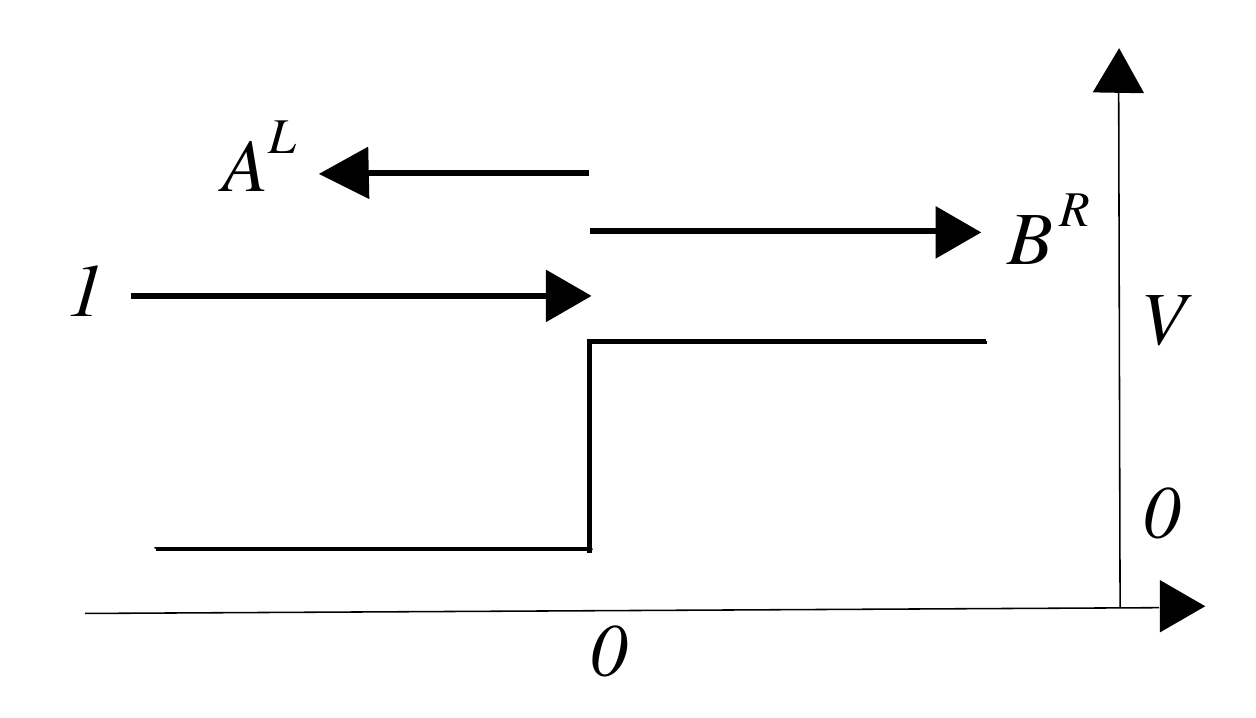}
\caption{The electron wave with unit amplitude, depicted as an arrow, is incident on the interface between semiconductors from the left. The reflected wave with amplitude $A^L$ and the transmitted wave with amplitude $B^R$ are also depicted. The interface between semiconductors is at the zero of $x$ axis. We count the potential energy from the bottom of the left semiconductor, the bottom of the right semiconductor has the potential $V$.}
\label{fig2}
\end{figure}

In light of this, we turn to the well-studied and widely applied Bastard conditions \cite{Bastard1, Bastard2}. We view these conditions as a long-wave limit of some Hamiltonian, akin to how the elastic springs equations \eqref{1longwave} represent the long-wavelength limit of lattice dynamics equations \eqref{1maineq}. Given our focus on long wavelengths, our study centers on the interface between semiconductors with simple zone structures. In such cases, electrons typically possess energies close to the bottom of a conduction band, corresponding to small wave vectors. Setting the origin of the energy axis at the bottom of the conduction band of the left crystal and denoting the energy of the bottom of the conduction band of the right crystal as $V$ (Fig. \ref{fig2}), the dispersion relations for the left and right semiconductors are:
\begin{align}
\varepsilon = \frac{p^{L^{2}}}{2m^{L}}, \, \, \,
\varepsilon = \frac{p^{R^{2}}}{2m^{R}}+V.
\end{align}
where $\varepsilon$ is the energy of an electron, $m^{L,R}$ and $p^{L,R}$ are effective masses and momenta of electrons on the left and the right sides, respectively.

Bastard conditions imply continuity of wave function at the interface and continuity of probability flow. If the wave of the unit amplitude is incident on the interface from the left, the equations are
\begin{align}
 1 + A^L = B^R 
\nonumber \\
v^L (1-A^L) = v^R B^R,
\label{3bastard}
\end{align} 
where $v=p/m$, the momenta of the electron divided by the effective mass of the electron. Analogous equations can be written for incidence from the other side. We solve both cases and find the amplitude matrix
\begin{align}
 \mathcal A = \begin{pmatrix} A^L & B^L \\ B^R & A^R \end{pmatrix} =
 \frac{1}{v^L + v^R} \begin{pmatrix} v^L - v^R & 2 v^R \\ 2 v^L & v^R -  v^L\end{pmatrix} .
\label{3solution}
\end{align} 
We do not introduce a special index that denotes amplitude matrix $\mathcal A$ for electron problem that would distinguish it from the phonon problem to keep formulas tidy. Given that the electron and phonon problems are discussed in different sections, we believe this approach will not lead to any confusion.

We see that the solution to the electron problem is structurally similar to the solution to the elastic springs problem \eqref{1elastic}. One can be obtained from the other by interchange $p/m \leftrightarrow v \rho $ with the corresponding side index. This gives us the idea that we can find an effective Hamiltonian for electron transmission with such an interchange. 

We introduce
\begin{align}
M^L_\pm = \mp i v^L - \beta
 \nonumber \\
M^R_\pm = \mp i v^R - \beta.
\label{1longwave}
\end{align}
 Where $\beta$ is some parameter with the dimension of velocity. Similarly
\begin{equation}
\mathcal M_\pm = \begin{pmatrix} M^L_\pm & \beta \\ \beta & M^R_\pm \end{pmatrix}, \,
\mathcal A = \begin{pmatrix} A^L & B^L \\ B^R & A^R \end{pmatrix}.
\end{equation} 
And the amplitude matrix
\begin{align}
\mathcal A = -\mathcal M_+^{-1}\, \mathcal M_-.
\end{align} 
It is easy to see that the solution is given by formula \eqref{3solution}. For matrices $\mathcal M_\pm$ to be a Hamiltonian, they should be multiplied on some parameter with the dimension of momenta. Note that we can straightforwardly find the amplitude matrix based on equation \eqref{3bastard}, but we believe that the listed approach is useful for it gives an insight into the possible form of the true Hamiltonian that describes the electron transfer through the interface.

We can also check the orthonormality of amplitude matrix \eqref{3solution}. The flow matrix in this case is given by
\begin{equation}
\mathcal F =  \begin{pmatrix} v^L & 0 \\ 0 & v^R \end{pmatrix},
\label{3elflow}
\end{equation} 

And, as it is easy to check, equation \eqref{1mainres} holds for the electronic amplitude matrix and flow matrix. Similarly to how we have done it in the previous section, we can introduce the unitary matrix $\mathcal U = \mathcal F^{1/2} \mathcal A \mathcal F^{-1/2}$ and write the relation between annihilation operators of the incident and departing electrons
\begin{align}
\begin{pmatrix} \hat c^L_{\leftarrow} \\ \hat c^R_{\rightarrow} \end{pmatrix} = \mathcal U \begin{pmatrix} \hat c^L_{\rightarrow } \\ \hat c^R_{\leftarrow} \end{pmatrix}.
\label{3connect}
\end{align} 
With this connection we can derive equations that connects distribution functions of electrons on both sides of the interface
\begin{align}
 f^L_{\leftarrow} = |A^L|^2 f^L_{\rightarrow} + \frac{v^L}{v^R} |B^L|^2 f^R_{\leftarrow}
\nonumber \\
f^R_{\rightarrow} = \frac{v^R}{v^L} |B^R|^2 f^L_{\rightarrow } + 
|A^R|^2 f^R_{\leftarrow}.
\label{3ElectronKinetic}
\end{align} 
In Ref. \cite{SemiKapitza} it was promised to derive matching equations for the distribution functions of electrons at the interface. The promise is kept here. To get to 3D equations considered in \cite{SemiKapitza} we should integrate over the angle of incidence and multiply on the density of states, that is proportional to momenta squared, hence we add multilier $p^2$. 

For simplification, we introduce the parameter $t$
\begin{align}
t=\frac{ v^Lv^R}{(v^L + v^R)^2}
\end{align} 
 and rewrite equations \eqref{3ElectronKinetic} as 
\begin{align}
 f^L_{\leftarrow} = (1-t) f^L_{\rightarrow} + t f^R_{\leftarrow}
\nonumber \\
f^R_{\rightarrow} = t f^L_{\rightarrow } + (1-t) f^R_{\leftarrow}.
\label{3GeneralKinetic}
\end{align} 
Since the sum of coefficients in each row is equal to one, we can naturally interpret $t$ as the electron transmission probability (similarly, we could have introduced phonon transmission probability at the end of Section 2).

In matrix notation we can rewrite this system of equations as
\begin{align}
\begin{pmatrix} f^L_{\leftarrow} \\ f^R_{\rightarrow} \end{pmatrix} = 
 \mathcal T \begin{pmatrix} f^L_{\rightarrow } \\ f^R_{\leftarrow} \end{pmatrix}.
\label{3Transf1}
\end{align} 
Elements of matrix $\mathcal T$ are obtained by taking a square of absolute values of elements of matrix $\mathcal U$. We can write it down compactly with the help of Hadamard product, denoted as "$\circ$"\, the elementwise matrix multiplication:
\begin{align}
 \mathcal T =  \overline {\mathcal U} \circ \mathcal U.
\label{3Hadamar}
\end{align}

Although the 3D transmission of electrons through the interface is explored in \cite{SemiKapitza}, we opt to consider the simpler 1D model presented here, utilizing the method introduced in \cite{Me}. This approach enables us to derive exact analytic expressions for kinetic coefficients that establish relationships between interfacial jumps (of temperature and electrochemical potential) and fluxes (heat and electric).

In the 3D case, obtaining the complete system of equations describing transport through the interface involves considering matching equations alongside Boltzmann equations for electrons on both sides of the interface and Chapman-Enskog conditions defining the temperature and concentration of electrons on both sides. This results in a set of integro-differential equations that can be challenging to solve.

However, in the 1D case, we can simplify the analysis. Following the method introduced in \cite{Me}, to define temperature and electrochemical potential on both sides of the interface, we augment the equations \eqref{3ElectronKinetic} with conditions:
\begin{align}
\frac{f_{\leftarrow}+f_{\rightarrow}}{2} = f_0.
\label{3Chapman}
\end{align} 
We omit the side index because these conditions are equal for both sides. Here $f_0$ denotes the Fermi distribution function
\begin{align}
f_0 = \frac{1}{\exp \left( \frac{\varepsilon - \zeta}{T}  \right) +1}.
\label{3Fermi}
\end{align} 
The temperatures $T^{L,R}$ and electrochemical potential $\zeta^{L,R}$ are assumed to be known for both sides. This conditions are just the definition of temperatures and electrochemical potentials on both sides. With such definition, we only have a system of linear algebraic equations to solve.

Expressions for the electric and heat flux are
\begin{align}
e \int d\varepsilon \left( f_{\rightarrow} - f_{\leftarrow} \right) = j
\nonumber \\
\int d\varepsilon \left( f_{\rightarrow} - f_{\leftarrow} \right) (\varepsilon - \zeta) = q
\label{3fluxes}
\end{align} 
We solve the system of equations (\ref{3GeneralKinetic}, \ref{3Chapman}, \ref{3Fermi}) and find $f^L_{\leftarrow}, f^L_{\rightarrow}, f^R_{\leftarrow}, f^R_{\leftarrow}$. We find that for both sides
\begin{align}
f_{\rightarrow} - f_{\leftarrow}  = \frac{t}{1-t} \left( f_0^L - f_0^R \right).
\label{3solution1}
\end{align} 
 We substitute this expression into \eqref{3fluxes} and obtain
\begin{align}
e \int d\varepsilon \frac{t}{1-t} \left( f_0^L - f_0^R \right) = j
\nonumber \\
\int d\varepsilon \frac{t}{1-t} \left( f_0^L - f_0^R \right) (\varepsilon - \zeta) = q
\label{3solution2}
\end{align} 

Now we can assume that the temperature and the electrochemical potential difference ($\Delta T$ and $\Delta \zeta$ respectively) on both sides are small compared to their absolute values. Then we can expand the Fermi distribution function to the Tailor series to the first order in these small parameters
\begin{align}
f_0^L =  f_0^R + \frac{\partial  f_0}{\partial T} \Delta T +  \frac{\partial  f_0}{\partial \zeta} \Delta \zeta
\end{align} 
Further, we introduce $x=\frac{\varepsilon - \zeta}{T}$ and rewrite the previous expression as
\begin{align}
f_0^L = f_0^R - \frac{\partial f_0}{\partial x} \frac{\varepsilon - \zeta}{T^2} \Delta T - \frac{\partial  f_0}{\partial x} \frac{1}{T} \Delta \zeta
\end{align} 
and substitute it into Eq. \eqref{3solution2}. We also introduce effective electric potential $\Delta U^* = \Delta \zeta/e$. We obtain 
\begin{align}
\left( -\frac{e}{T^2} \int d\varepsilon \frac{t}{1-t} \frac{\partial  f_0}{\partial x} (\varepsilon - \zeta) \right) \Delta T \hspace{14pt} + \hspace{14pt} \left( -\frac{e^2}{T} \int d\varepsilon \frac{t}{1-t} \frac{\partial  f_0}{\partial x} \right) \Delta U^* &= j
\nonumber \\
\left( -\frac{1}{T^2} \int d\varepsilon \frac{t}{1-t}\frac{\partial  f_0}{\partial x}  (\varepsilon - \zeta)^2 \right) \Delta T + 
\left( -\frac{e}{T} \int d\varepsilon \frac{t}{1-t}\frac{\partial  f_0}{\partial x}  (\varepsilon - \zeta) \right) \Delta U^* &= q
\label{3final}
\end{align} 
Expressions in the circular brackets are relations between interfacial jumps and fluxes.

 We see, that at $t=1$ we have infinite values of these coefficients, indicating the absence of jumps in this case.  This aligns with the expected behavior, as $t=1$ corresponds to a crystal plane in a uniform lattice. A reasonable model should not predict any jumps in a homogeneous medium. Additionally, the Onsager reciprocity holds, as the symmetrical coefficients (between $\Delta T$ and $j$ and between $\Delta U^*$ and $q$) only differ in the multiplier $T$.

This demonstrates that the provided toy model, despite its simplicity, offers the correct qualitative description of electron transport through the interface. A more precise numerical description is presented in Ref. \cite{SemiKapitza}.

%%%%%%%%%%%%%%%%%%%%%%%%%%

\section{Irreversability of transport through the interface}

Now we look again at equations \eqref{3connect}. We can invert them, and since matrix invert for a unitary matrix is its Hermitian conjugate, we get
\begin{align}
\begin{pmatrix} \hat c^L_{\rightarrow } \\ \hat c^R_{\leftarrow} \end{pmatrix} =
\mathcal U^\dag\begin{pmatrix} \hat c^L_{\leftarrow} \\ \hat c^R_{\rightarrow} \end{pmatrix} .
\label{4inverse}
\end{align} 
With previously established procedure we can derive matching equations for the distribution functions and we get
\begin{align}
 f^L_{\rightarrow} = (1-t) f^L_{\leftarrow} + t f^R_{\rightarrow}
\nonumber \\
f^R_{\leftarrow} = t f^L_{\leftarrow} + (1-t) f^R_{\rightarrow}.
\label{4ReverseKinetic}
\end{align} 
Again we accomponie this with additional conditions \eqref{3Chapman} and solve it. We find 
\begin{align}
f_{\rightarrow} - f_{\leftarrow}  = - \frac{t}{1-t} \left( f_0^L - f_0^R \right) 
\end{align} 
We compare it with expression \eqref{3solution1} and observe that we have obtained a minus sign before the expression, resulting in a change in the sign of all fluxes.

Let's reflect on the implications of this change. Equations \eqref{3connect} represent the transformation from the previous state (incident electrons) to the next state (departing electrons). Conversely, equations \eqref{4inverse} portray transformations from the next state to the previous state. The inversion, in essence, signifies a reversal in the direction of time. Consequently, with this time inversion, heat flows from the colder region to the hotter region, aligning with the time asymmetry observed in thermodynamics.

We can consider it from the other viewpoint. Let us try to invert the equation \eqref{3GeneralKinetic}. We obtain
\begin{align}
 f^L_{\rightarrow} = \frac{1-t}{1-2t} f^L_{\leftarrow} - \frac{t}{1-2t} f^R_{\rightarrow}
\nonumber \\
f^R_{\leftarrow} = -\frac{t}{1-2t} f^L_{\leftarrow} + \frac{1-t}{1-2t}f^R_{\rightarrow}.
\label{4ReverseAlternative}
\end{align} 
It does not coincide with equation \eqref{4ReverseKinetic}! The coefficients can no longer be interpreted as probabilities. While the sum of the coefficients in one row still equals 1, either one of them is negative, or $t=1/2$, and the matrix $\mathcal T$ is not invertible. Thus, we have transitioned from reversible wave equations to irreversible kinetic equations. We need to understand this shift both mathematically and physically.

From a mathematical perspective, matrix $\mathcal T$ belongs to the class of bistochastic matrices. These are square matrices of nonnegative real numbers, each of whose rows and columns sum to one. This can be directly observed for the two-dimensional case, and for every dimension, it follows from formula \eqref{3Hadamar} and the property of a unitary matrix that the sum of the squares of absolute values of elements in every row and every column is equal to one.

Bistochastic matrices, in general, lack inverses within the same class of matrices. While every square matrix with a nonzero determinant has a matrix inverse, a bistochastic matrix's inverse will not be a bistochastic matrix. The only exception to this rule is permutation matrices, which exclusively consist of ones and zeroes. This can be directly observed in the case of a two-dimensional bistochastic matrix (Eq. \ref{4ReverseAlternative}), and a general proof can be easily provided.

Let's compare it to the case of unitary matrices. Unitary matrices form a group, implying that the multiplication of unitary matrices yields another unitary matrix. This group exhibits associativity in multiplication, contains unity (the identity matrix), and every element has an inverse. Notably, the inverse of a unitary matrix is also a unitary matrix.

Bistochastic matrices, when multiplied, also result in a bistochastic matrix, a well-known property extending from the broader category of stochastic matrices \cite{FTMC}. Multiplication is associative since matrix multiplication follows this property, and the identity matrix is bistochastic, establishing the set as a monoid. While there are generally no inverses in the set of bistochastic matrices, this algebraic property aligns with the physical property of irreversibility inherent in processes described by bistochastic matrices.

The subset of two-by-two bistochastic matrices possesses particularly elegant characteristics. The coefficients are so constrained that any two-dimensional bistochastic matrix can be uniquely characterized by a single parameter. Specifically, if we set one element to $t$, the element in the same row and the element in the same column become $1-t$, and these conditions dictate that the cross-diagonal element equals $t$. This allows us to introduce a remarkably simple rule for the multiplication of two-by-two bistochastic matrices, employing an alternative parameterization:
\begin{align}
\mathcal T_x =
\frac{1}{2} \begin{pmatrix} 1 & 1 \\ 1 & 1 \end{pmatrix} +
\frac{1}{2} \begin{pmatrix} 1 & -1 \\ -1 & 1 \end{pmatrix} x  .
\label{4parameters}
\end{align} 
where $x$ belongs to the interval $[-1, 1]$. We can see that
\begin{align}
\frac{1}{2} \begin{pmatrix} 1 & 1 \\ 1 & 1 \end{pmatrix} * \frac{1}{2} \begin{pmatrix} 1 & 1 \\ 1 & 1 \end{pmatrix} = \frac{1}{2} \begin{pmatrix} 1 & 1 \\ 1 & 1 \end{pmatrix},
\nonumber \\
\frac{1}{2} \begin{pmatrix} 1 & -1 \\ -1 & 1 \end{pmatrix} * \frac{1}{2} \begin{pmatrix} 1 & -1 \\ -1 & 1 \end{pmatrix} = \frac{1}{2} \begin{pmatrix} 1 & -1 \\ -1 & 1 \end{pmatrix}
\nonumber \\
\begin{pmatrix} 1 & -1 \\ -1 & 1 \end{pmatrix} * \begin{pmatrix} 1 & 1 \\ 1 & 1 \end{pmatrix} = 0,
\, \, \,
\begin{pmatrix} 1 & 1 \\ 1 & 1 \end{pmatrix} * \begin{pmatrix} 1 & -1 \\ -1 & 1 \end{pmatrix} = 0,
\label{4parameters}
\end{align} 
So 
\begin{align}
\mathcal T_{x_1} \mathcal T_{x_2} =
\frac{1}{2} \begin{pmatrix} 1 & 1 \\ 1 & 1 \end{pmatrix} +
\frac{1}{2} \begin{pmatrix} 1 & -1 \\ -1 & 1 \end{pmatrix} x_1 x_2 
= \mathcal T_{x_1 x_2}.
\label{4parameters}
\end{align} 
Matrices multiply like numbers on the  $[-1, 1]$ segment. The absence of inverses for elements other than $x=1, x=-1$ is especially obvious from this formula. The matrices for the case of $x=1, x=-1$ are
\begin{align}
\mathcal T_{x=1} = \begin{pmatrix} 1 & 0\\ 0 & 1 \end{pmatrix}, \, \, \,
 \mathcal T_{x=-1} = \begin{pmatrix} 0 & 1 \\ 1 & 0 \end{pmatrix}.
\label{4cases}
\end{align} 
Theses are permutation matrices, which illustrates more general n-dimensional case. 

With formula \eqref{4parameters} we can make an observation that for any matrix other than $x=1, x=-1$ its repeated multiplication upon itself has a certain limit
\begin{align}
\mathcal T^{\infty} = \mathcal T_{x=0} =
\frac{1}{2} \begin{pmatrix} 1 & 1 \\ 1 & 1 \end{pmatrix}.
\label{4Markov}
\end{align} 
It is a simple special case of the famous fundamental theorem of Markov chains \cite{FTMC}.

Now we want to understand what physical conditions leeds to the observed irreversability. Consider a classical field where plane waves of this field are incident on the interface from both sides. For concreteness, let's envision oscillations of a crystalline lattice, an example elucidated in Section 2. However, the forthcoming reasoning is versatile and can be applied to any classical field. Matrix $\mathcal A$ was derived as a combination of solutions for two cases: the incidence of a wave with unit amplitude from the left side and from the right side. In a more general scenario, the amplitudes of incident waves are denoted as $A^L_{\rightarrow}$ and $A^R_{\leftarrow}$, while the amplitudes of departing waves are $A^L_{\leftarrow}$ and $A^R_{\rightarrow}$. By linearity, they are interconnected by matrix $\mathcal A$:
\begin{align}
\begin{pmatrix}  A^L_{\leftarrow} \\  A^R_{\rightarrow} \end{pmatrix} = 
\mathcal A \begin{pmatrix}  A^L_{\rightarrow } \\  A^R_{\leftarrow} \end{pmatrix} 
\label{4amps}
\end{align} 
We introduce normalized amplitudes
\begin{align}
\begin{pmatrix}  a^L_{\leftarrow} \\  a^R_{\rightarrow} \end{pmatrix} = \mathcal N^{-1} 
\begin{pmatrix}  A^L_{\leftarrow} \\  A^R_{\rightarrow} \end{pmatrix}, \, \, \,
 \begin{pmatrix}  a^L_{\rightarrow } \\  a^R_{\leftarrow} \end{pmatrix} =
\mathcal N^{-1} \begin{pmatrix}  A^L_{\rightarrow } \\  A^R_{\leftarrow} \end{pmatrix}.
\label{4normal}
\end{align} 
For normalized amplitudes we have
\begin{align}
\begin{pmatrix}  a^L_{\leftarrow} \\  a^R_{\rightarrow} \end{pmatrix} = 
\mathcal U \begin{pmatrix}  a^L_{\rightarrow } \\  a^R_{\leftarrow} \end{pmatrix}.
\label{4namps}
\end{align} 
Where we have used equality \eqref{2transform}. That is a reversible transformation and we can write down 
\begin{align}
\begin{pmatrix}  a^L_{\rightarrow } \\  a^R_{\leftarrow} \end{pmatrix} = 
\mathcal U^\dag \begin{pmatrix}  a^L_{\leftarrow} \\  a^R_{\rightarrow} \end{pmatrix}.
\label{4reverse}
\end{align} 
Indeed $\mathcal U^\dag$ is inverse to $\mathcal U$ and is also unitary.

Now amplitudes are complex numbers with given absolute values and the argument $a = |a|e^{i\phi}$. We want to calculate the intensity which is a real, observable parameter. We write down equations \eqref{4namps} componentwise and have
\begin{align}
a^L_{\leftarrow} = U_{11} a^L_{\rightarrow} + U_{12} a^R_{\leftarrow}
\nonumber \\ 
a^R_{\rightarrow} = U_{21} a^L_{\rightarrow} + U_{22} a^R_{\leftarrow}.
\label{4components}
\end{align} 
We multiply every part on its complex conjugate and obtain
\begin{align}
|a^L_{\leftarrow}|^2 = |U_{11}|^2 |a^L_{\rightarrow}|^2 + |U_{12}|^2
|a^R_{\leftarrow}|^2 +  \overline U_{11} U_{12}  \overline a^L_{\rightarrow} a^R_{\leftarrow} + U_{11}  \overline U_{12} a^L_{\rightarrow}  \overline a^R_{\leftarrow}
\nonumber \\ 
|a^R_{\rightarrow}|^2 = |U_{21}|^2 |a^L_{\rightarrow}|^2 + |U_{22}|^2
|a^R_{\leftarrow}|^2 +  \overline U_{21} U_{22}  \overline a^L_{\rightarrow} a^R_{\leftarrow} + U_{21}  \overline U_{22} a^L_{\rightarrow}  \overline a^R_{\leftarrow}
\label{4squares}
\end{align}
Let us denote the difference between the arguments of $a^L_{\rightarrow}$ and $a^R_{\leftarrow}$ as 
$\Delta \phi$. We also introduce the intensities $I = |a|^2$. We get 
\begin{align}
I^L_{\leftarrow} = |U_{11}|^2 I^L_{\rightarrow} + |U_{12}|^2 I^R_{\leftarrow} + 
\overline U_{11} U_{12} |a^L_{\rightarrow} a^R_{\leftarrow}| e^{i\phi} + 
U_{11}  \overline U_{12} |a^L_{\rightarrow} a^R_{\leftarrow}| e^{- i\phi}
\nonumber \\ 
I^R_{\rightarrow}= |U_{21}|^2 I^L_{\rightarrow} + |U_{22}|^2 I^R_{\leftarrow} +
 \overline U_{21} U_{22} |a^L_{\rightarrow} a^R_{\leftarrow}| e^{i\phi} + 
U_{21}  \overline U_{22} |a^L_{\rightarrow} a^R_{\leftarrow}|  e^{- i\phi}
\label{4intensity}
\end{align} 
So if there is a fixed $\Delta \phi$ we can deduce intensities of transmission from the complex amplitudes, and the amplitude transformation is given by equation \eqref{4amps}.

We can sum both equations in the system \eqref{4intensity} and get
\begin{align}
I^L_{\leftarrow} + I^R_{\rightarrow} =  I^L_{\rightarrow} + I^R_{\leftarrow},
\label{4full}
\end{align} 
because $|U_{11}|^2 + |U_{21}|^2 = 1$, $|U_{12}|^2 + |U_{22}|^2 = 1$ and $\overline U_{11} U_{12} +  \overline U_{21} U_{22} = 0$, which is the consequence of orthonormality of columns in a unitary matrix. So we see that the total energy flux is conserved.

Here we uncover the physical meaning of relation \eqref{1flowort} and, more broadly, the "orthonormality of flow" condition \eqref{1mainres}. This condition ensures that the interference of waves incident on the interface will not violate the conservation of energy. As energy is conserved in any physically meaningful system, this relation holds for any field. It guarantees that we can construct a unitary matrix $\mathcal U$ from the amplitude matrix $\mathcal A$ using the formula \eqref{2transform}. In turn, this ensures that we can quantize the field using the arguments presented in Section 3 and subsequently derive a kinetic equation from the quantized equation.

Returning to classical fields, let's consider incoherent waves. This implies that we "do not know" the complex amplitudes, only the intensities of the waves. In other words, we "do not know" the arguments of the amplitudes in Eq. \eqref{4intensity}, only their absolute values. To determine how the intensities transform at the interface, as commonly done with incoherent oscillations \cite{Optics}, we average over different values of arguments. This entails integrating equations \eqref{4intensity} over the range from $0$ to $2\pi$ and dividing by $2\pi$. The result is:
\begin{align}
I^L_{\leftarrow} = |U_{11}|^2 I^L_{\rightarrow} + |U_{12}|^2 I^R_{\leftarrow} 
\nonumber \\ 
I^R_{\rightarrow}= |U_{21}|^2 I^L_{\rightarrow} + |U_{22}|^2 I^R_{\leftarrow}.
\label{4incoherent}
\end{align} 
In matrix form it is
\begin{align}
\begin{pmatrix} I^L_{\leftarrow} \\ I^R_{\rightarrow} \end{pmatrix} = 
 \mathcal T \begin{pmatrix} I^L_{\rightarrow } \\ I^R_{\leftarrow} \end{pmatrix}.
\label{4Transf}
\end{align} 
where $\mathcal T$ is given by the formula \eqref{3Hadamar}. We see that intensities of incoherent waves transforms at the interface the same way as distribution function of particles \eqref{3Transf1}.

Let us go back to the formulae (\ref{2Kronecker}, \ref{2NumbOps}). Let us suppose that the state of departing particles is given by the superposition of states with the various ocupation numbers
$|D\rangle = A_1 |n_1, n_2\rangle + A_2 |n_1 + 1, n_2 - 1\rangle $. Then
\begin{align}
\langle D| \hat a_1^\dag \hat a_2 |D\rangle = \overline A_2 A_1 \langle n_1 + 1, n_2 - 1| \hat a_1^\dag \hat a_2 |n_1, n_2\rangle =  \overline A_2 A_1 \sqrt{(n_1 + 1) n_2} \neq 0
\label{4nosuper}
\end{align} 
Now if we apply the incoherence condition, like we did to get to equation \eqref{4incoherent}, that is, we integrate over the phase difference between $A_2$ and $A_1$, we get $\langle D| \hat a_1^\dag \hat a_2 |D\rangle = 0$. And this gives us the condition \eqref{2Kronecker} which is needed to get to the kinetic equation \eqref{2PhononKinetic}. 

In Eq. \eqref{2PhononKinetic} there is still a probability, the parameter $t$ denotes the probability for the particle to transmit through the interface. But there is no superposition: the state of departing particles can have different occupation numbers with different probabilities, but it can not be in a superposition of states with different occupation numbers. This is in complete agreement with the recent observation that quantum entanglement implies coherence \cite{Entanglement}. The incoherence condition imposed on a quantum system means exactly the absence of superposition. So incoherent flow of quantum particles has probabilistic properties but does not have wave properties, which are interference, in the classical picture or superposition in the quantum picture. And such incoherent flow is described by kinetic equations, which are irreversible.  

We arrive at the following conclusions:
\begin{itemize}
\item Incoherence equals irreversibility.
\item The flow of incoherent waves behaves like the flow of particles.
\end{itemize}

In the next section, we will present a simple experiment, that can directly verify the first conclusion. The experimental observation of interfacial kinetic coefficients for electron gas, which we have talked about in the Section 4 would be a serious confirmation of the second conclusion.

%%%%%%%%%%%%%%%%%%%

\section{Light transport through the interface, coherence and reversability}
In the previous section, we established that imposing the incoherence condition on the wave equations \eqref{4namps} leads to irreversible equations. However, the equation \eqref{4namps} is equivalent to the equation \eqref{4reverse}. And if we imply the incoherence condition to Eq. \eqref{4reverse} we get the reverse arrow of time, like in the case of equations (\ref{4inverse}, \ref{4ReverseKinetic}). In principle, we can use the equation 
\begin{align}
\mathcal U^{-1/3}\begin{pmatrix}  a^L_{\leftarrow} \\  a^R_{\rightarrow} \end{pmatrix} = 
\mathcal U^{2/3} \begin{pmatrix}  a^L_{\rightarrow } \\  a^R_{\leftarrow} \end{pmatrix}.
\label{5example}
\end{align}
which is equivalent to equations (\ref{4namps}, \ref{4reverse}) and obtain a kinetic equation from there. How do we choose the right way to get a kinetic equation with the right direction of time?

Equation \eqref{4namps}, like equation \eqref{1maineq}, corresponds to one incident wave being separated by the interface into two distinct departing waves. Conversely, equation \eqref{4reverse} corresponds to two incident waves being assembled into one departing wave. Formally, waves that are incident on the interface from different directions cannot be causally related. Therefore, for the right form of the equation, we should keep it in the form where the future state of the system (departing waves) is a transformation of the previous state (incident waves).

Intuitively, we know that the first type of process should describe physical reality, akin to the burning paper transforming into smoke, which is a real process, while smoke assembling into the clean paper is a time-reversed process. This aligns with our intuition for entropy growth in real-world situations. However, it is not the way coherent waves behave. Two separate coherent waves can be assembled into one wave.

Let us consider the classical topic of transmission of light through the interface. The amplitudes of transmitted and reflected waves for this case are given by the famous Fresnel formula. For simplicity, we would talk about the interface between the air and some non-magnetic material, with refraction index $n$. The air side would be the left side and the material side would be the right side. Also, we would consider the waves to be linearly polarized and the direction of polarization to be perpendicular to the plane of incidence. In this case, the amplitude matrix is given by 
\begin{align}
 \mathcal A = \setstretch{1.5}
 \frac{1}{\sin(\phi+\psi)} \begin{pmatrix} \sin(\psi-\phi) & 2 \sin\psi\cos\phi \\2 \sin\phi\cos\psi &  \sin(\phi-\psi) \end{pmatrix} .
\label{5Fresnel}
\end{align} 
Where angles $\phi$ and $\psi$ are angles between the propagation direction of the wave and the axis perpendicular to the interface, $\phi$ is the angle on the air side and $\psi$ is the angle on the material side. By Snell's law $\sin\phi$ = $n \sin\psi$. 

The energy flow is the Poynting vector. Conserved is only the component that is perpendicular to the interface. We can easily see that the components of flow parallel to the interface should not be equal on both sides of the interface, since in the case of full reflection on one side there is no energy flow and on the other side there is.

In the considered case the intensity in the direction perpendicular to the interface is proportional to the velocity perpendicular to the interface and to $n^2$ so the flow matrix is given by
\begin{align}
\mathcal F = \begin{pmatrix} \cos \phi & 0 \\ 0 & n \cos \psi \end{pmatrix}  = 
\begin{pmatrix} \cos \phi & 0 \\ 0 & \frac{\sin\phi \cos\psi}{\sin\psi}  \end{pmatrix}.
\label{5flow}
\end{align} 
We have used Snell's law for the second equality.

With this flow matrix and amplitude matrix, we can find the unitary transform matrix. We substitute relations (\ref{5Fresnel}, \ref{5flow}) to equation \eqref{2transform} and obtain
\begin{align}
 \mathcal U = \setstretch{1.5}
 \frac{1}{\sin(\phi+\psi)} \begin{pmatrix} \sin(\psi-\phi) & 2 \sqrt{\sin\psi\cos\phi\sin\psi\cos\phi}
\\ 2  \sqrt{\sin\psi\cos\phi\sin\psi\cos\phi} &  \sin(\phi-\psi) \end{pmatrix}.
\label{5unitary}
\end{align} 

The first column of the amplitude matrix consists of amplitudes of reflected and transmitted waves in the case of the incidence of the wave of the unit amplitude from the air side, and the second is amplitudes in the case of the incidence from the material side. We can write it as 
\begin{align}
\begin{pmatrix} A^L & B^L \\ B^R & A^R  \end{pmatrix} = \mathcal A \begin{pmatrix} 1 & 0 \\ 0 & 1 \end{pmatrix}. 
\label{5forward}
\end{align} 
We want to find the amplitudes of the waves coming from both sides so that they will assemble  into a single wave. That means we want to substitute the matrix on the right side instead of the identity matrix so that on the left side we get an identity matrix instead of the matrix of amplitudes. As it is easy to see, the correct matrix is $\mathcal A^{-1}$.

\begin{figure}
\includegraphics[width=0.88\textwidth]{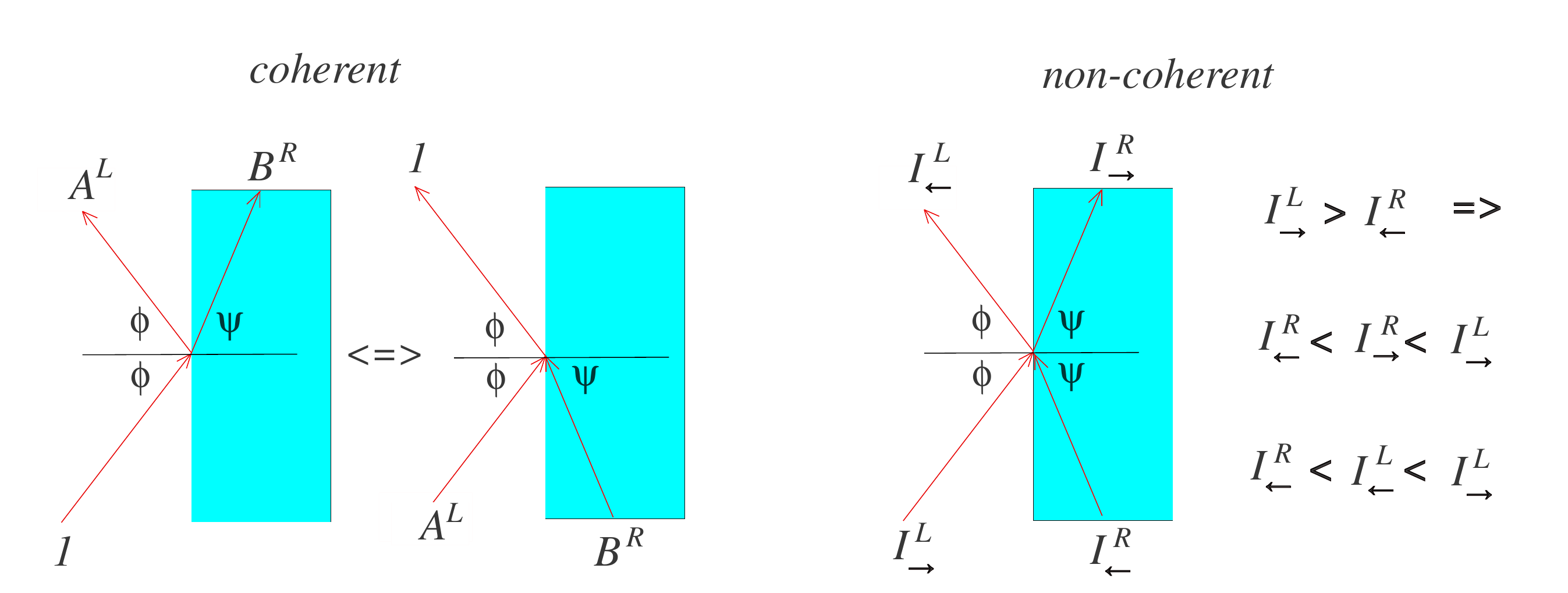}
\caption{In the first segment of the image, labeled "\textit{coherent}," a laser beam with unit amplitude is incident on the interface between air and a transparent material with a refraction index $n$ (depicted in light blue). The angle between the beam and the line perpendicular to the air-side interface is $\phi$, and on the opposite side, it is $\psi$. If the beam separates into two beams with amplitudes $A^L$ and $B^R$, then the two beams, both with amplitudes $A^L$ and $B^R$ and the same phase, combine to form a single beam with unit amplitude.\\
In the second segment of the image, labeled "\textit{non-coherent}," the same experimental configuration is reproduced for non-coherent light. It is demonstrated that two beams of incoherent light exhibit fundamentally different behavior. It is impossible to achieve a departing light intensity on any side greater than the larger of the two intensities of the incident beams. }
\label{Fassemble}
\end{figure}

We can quickly find that $\mathcal A = \mathcal A^{-1}$ with the help of the following line
\begin{align}
\mathcal A =  \mathcal N \mathcal U \mathcal N^{-1} => \mathcal A^{-1} =  \mathcal N \mathcal U^\dag \mathcal N^{-1} \nonumber \\
\mathcal U^\dag = \mathcal U => \mathcal A^{-1} = \mathcal A.
\end{align} 
Or we can simply verify that $\mathcal A^2 = \mathcal E$ by direct computation.

That implies, when combining beams with amplitudes $A^L, B^R$ at the angles $\phi, \psi$ at one point of the interface they will assemble into one beam with a unit amplitude on the side of the air, and if we combine beams with amplitudes $A^R, B^L$ we will get one ray with a unit amplitude on the side of the material (Fig. \ref{Fassemble}).

For incoherent radiation the intensities transform by transmission matrix $\mathcal T$ \eqref{4Transf}. We substitute the unitary matrix given by formula \eqref{5unitary} to equation \eqref{3Hadamar} and get
\begin{align}
 \mathcal T = \setstretch{1.5}
\frac{1}{\sin^2(\phi+\psi)} \begin{pmatrix} \sin^2(\psi-\phi) & 4\sin\psi\cos\phi \sin\phi\cos\psi
\\ 4 \sin\psi\cos\phi\sin\phi\cos\psi &  \sin^2(\phi-\psi) \end{pmatrix}.
\label{5transmission}
\end{align} 

We can observe that incoherent light behaves principialy different then the coherent one. Let $ I^L_{\rightarrow }> I^R_{\leftarrow}$. We can observe that $ I^L_{\rightarrow }> I^L_{\leftarrow}> I^R_{\leftarrow}$, $ I^L_{\rightarrow }> I^R_{\rightarrow}> I^R_{\leftarrow}$. Indeed
\begin{align}
I^L_{\leftarrow} = (1-t) I^L_{\rightarrow } + t I^R_{\leftarrow} => \, \, \,
I^L_{\rightarrow } - I^L_{\leftarrow} = t ( I^L_{\rightarrow } - I^R_{\leftarrow})>0 =>
I^L_{\rightarrow } > I^L_{\leftarrow}.
\end{align} 
An analogous proof can be given for three other inequalities. We can not get the intensity of the departing light on one side to be larger than the largest of the two intensities of the incident beams (Fig. \ref{Fassemble}). 

\begin{figure}
\includegraphics[width=0.94\textwidth]{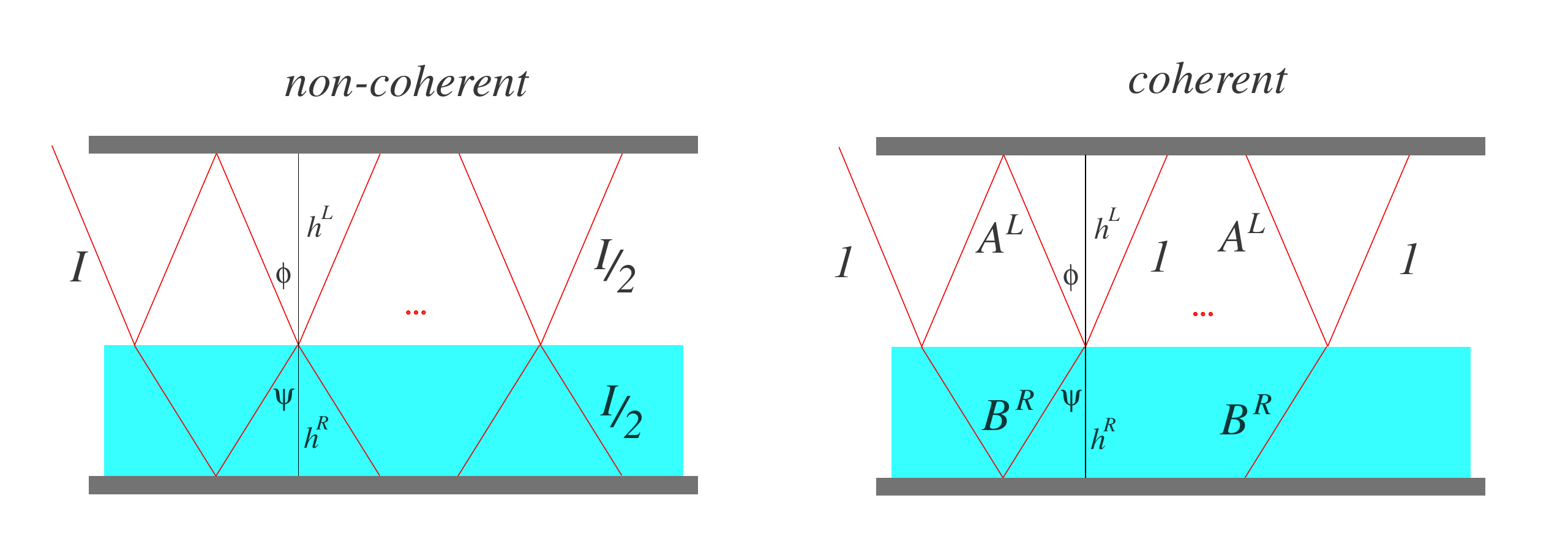}
\caption{In the first segment of the image, labeled "\textit{non-coherent}," a non-coherent beam of light with intensity $I$ is incident on the interface between air and a transparent material with a refraction index $n$ (depicted in light blue). The beam traverses the interface, splits into two beams, reflects in mirrors (depicted in dark grey), and subsequently meets again at the interface. This recurrence is ensured by the angles $\phi$ and $\psi$ and the distances between the interface and mirrors satisfying the relation $h^L/h^R = \tan \psi/\tan \phi$. The meeting beams exchange intensities, reflect once more, reconvene at the interface again, and repeat this process multiple times. If this occurs enough times, the intensities of the beams on both sides of the interface equalize to $I/2$.\\
In the second segment of the image, labeled "\textit{coherent}," the same experimental configuration is reproduced for the laser. The laser beam, initially split by the interface, repeatedly assembles into one beam, as long as the coherence length of the laser is sufficient.}
\label{Fequilibrium}
\end{figure}

We aim to examine the asymptotic behavior of a beam traversing an interface, undergoing a bifurcation into two beams. Subsequently, these beams reflect, reconvene at the interface, interchange intensities, reflect once more, reconvene at the interface again, and iterate through this process multiple times (Fig. \ref{Fequilibrium}). According to geometric optics principles, the distance between the interface and the mirrors on both sides ($h^L, h^R$) must satisfy the relation $h^L/h^R = \tan \psi/\tan \phi$ to ensure the infinite recurrence of the two beams at the interface.

Since at every instance of beams meeting at the interface their intensities transform by the formula \eqref{4Transf}, the intensities after $n$ meets will be given by 
\begin{align}
\begin{pmatrix} I^L_{\leftarrow} \\ I^R_{\rightarrow} \end{pmatrix}_n = 
 \mathcal T^n \begin{pmatrix} I^L_{\rightarrow } \\ I^R_{\leftarrow} \end{pmatrix}_0.
\label{5nTransf}
\end{align} 
the columns on the right side represent intensities at the first meet of beams. Since for matrices, $\mathcal T$ exists a limiting behavior described by the  formula \eqref{4Markov}, given enough meets of beams at the interface, the intensities on both sides will become
\begin{align}
\begin{pmatrix} I^L_{\leftarrow} \\ I^R_{\rightarrow} \end{pmatrix}_{\infty} =
\frac{1}{2} \begin{pmatrix} 1 & 1 \\ 1 & 1 \end{pmatrix} \begin{pmatrix} I^L_{\rightarrow } \\ I^R_{\leftarrow} \end{pmatrix}_0,
\label{5Limit}
\end{align} 
where the limiting in intensities are denoted by index $\infty$. But that means simply that 
\begin{align}
 I^L_{\leftarrow \infty} = I^R_{\rightarrow \infty} = (I^L_{\rightarrow 0} + I^R_{\leftarrow 0})/2 = I_{full}/2.
\label{5Limit2}
\end{align} 
Regardless of the initial side on which the first beam is incident, after a sufficient number of meetings and reflections, the intensities on both sides will equalize. The sum of these intensities will be half of the original intensity of the first beam incident on the interface.

On the contrary, coherent light does not exhibit any limiting behavior. As demonstrated, the entirety of the intensity from two intersecting rays can be directed into a single direction. In the geometry of the experiment described above, where beams meet repeatedly after reflections, starting with a single beam leads to a process of splitting into two, merging back into one, splitting again, and so forth, as long as the coherence length of the laser is sufficient (refer to Fig. \ref{Fequilibrium}). This outcome arises from the equality $\mathcal A^2 = 1$, signifying that performing the same transformation twice yields the initial result. Additionally, the fact that two beams meet again at the interface ensures that the phase shift is identical for both beams.

It is interesting to note that in statistical physics, the conventional perspective often posits irreversibility as a consequence of entropy increase. However, this viewpoint faces criticism, notably in classical texts such as Landau and Lifshitz \cite{LdStat}. In contrast, we propose that irreversibility stems from the incoherent evolution of the system. To clarify, the increase in entropy is a consequence of irreversibility at a microscopic level. Our kinetic problem, formulated and explored in Section 4, involves two sets of equations: the matching equations of distribution functions at the interface \eqref{3GeneralKinetic} and thermalization equations \eqref{3Chapman}. While the latter allows us to define the temperature and other thermodynamic parameters of the system, it is the matching equations that, as we have demonstrated mathematically, give rise to irreversibility. In this section, we present an example of a system (crossing beams at the interface) not described in terms of temperatures but still exhibiting irreversible behavior.

The experimental validation of the aforementioned theory regarding light transmission through the interface would inevitably lead to the conclusion that coherent and incoherent light exhibit fundamentally distinct behaviors. Specifically, it would establish that the transmission of coherent light through the interface is reversible, whereas the transmission of incoherent light is irreversible. Such experimental results would strongly support the perspective presented in this manuscript, asserting that incoherence is the underlying source of irreversibility.

\section{Summary and conclusions}

We have investigated the transmission of phonons, electrons, and photons through the interface between media. We have found that the conservation of energy law in the interfacial case manifests itself as the conservation of the energy flow vector that is perpendicular to the interface. In case two coherent waves are incident on the interface the energy flow is conserved for any combination of amplitudes and phases incident waves have. This fact is expressed in equation \eqref{1mainres}. 

When elastic waves are incident on the interface, the amplitudes of waves transform with matrix \eqref{1elastic} and the flow matrix is given by expression \eqref{1flowmat}. For electrons incident on the interface of a semiconductor, their amplitudes may be found with Bastard conditions, and they transform with matrix \eqref{3solution} and the flow matrix is given by \eqref{3elflow}.  In the case of light propagation through an interface between media with different refractive indices, the amplitude transformation is governed by the matrix \eqref{5Fresnel}, and the flow matrix is defined by \eqref{5flow}. In all three cases, the equation \eqref{1mainres} remains valid.

We impose the condition of non-coherency on the transformation of amplitudes (see Eqs. \ref{4incoherent}, \ref{4Transf} and the text around). With this condition transformation of intensities at the interface is given by bistochastic matrices, which are obtained from the derived unitary matrices by formula \eqref{3Hadamar}. This family of matrices does not have matrix inverses (inside the family). Physically it means that processes described by such matrices are irreversible. We conclude that the transmission of non-coherent waves through the interface is irreversible. We suggest two experiments that highlight the principal differences between the behavior of coherent and non-coherent light at the interface (Figs. \ref{Fassemble}, \ref{Fequilibrium}) and confirm the described hypothesis.

It is the consequence of equation \eqref{1mainres} that we can form a unitary matrix by relation $\mathcal U = \mathcal F^{-1/2} \mathcal A \mathcal F^{1/2}$ (see Eq. \ref{2transform} and the text around). According to general principles of quantum mechanics, the time evolution of a quantum system is described by unitary transformation. We associate the constructed unitary matrix with the transition from the state of quantum particles incident on the interface to the state of departing particles.  

From a quantum theory point of view, the non-coherence condition means no-superposition condition \eqref{4nosuper}. We apply it to the unitary matrix of transformation of amplitude operators and derive the equation of transformation of the distribution functions of particles at the interface, the matching equations (\ref{2PhononKinetic}, \ref{3Transf1}). The distribution functions of particles are transformed by bistochastic matrices like intensities of non-coherent waves. So the matching equations are time-irreversible. It is the desired property because kinetic equations should correctly describe the irreversible process of heat transport and other transport phenomena. We conclude that kinetic equations are obtained from quantum-mechanical equations by imposing non-coherence conditions on them.

\acknowledgments{ The author is grateful for A.\,Ya. Vul and C. Pastorino for their attention to the presented investigation.}

\end{document}